\documentclass[10pt]{article}

\usepackage{amsmath,amssymb,dsfont,slashed,color,hyperref}
\usepackage{mathtools}
\usepackage{authblk}
\usepackage[toc,page]{appendix}
\usepackage{makecell}

\definecolor{armygreen}{rgb}{0.29, 0.33, 0.13}

\newcommand{\half}{\frac{1}{2}}

\newcommand{\JJ}{\mathds{J}}
\newcommand{\KK}{\mathds{K}}
\newcommand{\DD}{\mathds{D}}
\newcommand{\TT}{\mathds{T}}
\newcommand{\VV}{\mathds{V}}
\newcommand{\ZZ}{\mathds{Z}}
\newcommand{\GG}{\mathds{G}}

\newcommand{\CC}{\mathds{C}}
\newcommand{\QQ}{\mathds{Q}}
\newcommand{\QQb}{\overline{\mathds{Q}}}

\newcommand{\uantof}{Departamento de Física, Universidad de Antofagasta, Aptdo. 02800, Chile}

\newcommand{\cecs}{Centro de Estudios Cient\'{\i}ficos (CECs), Arturo Prat 514, Valdivia, Chile}

\begin{document}

\title{Embedding of rank two supercharges in the superconformal algebra}
\author[1]{P. D. Alvarez \thanks{E-mail: \href{mailto:pedro.alvarez@uantof.cl}{\nolinkurl{pedro.alvarez@uantof.cl}}}}

\author[1]{R. A. Chavez \thanks{E-mail: \href{mailto:rafael.chavez.linares@uantof.cl}{\nolinkurl{rafael.chavez.linares@uantof.cl}}}}
\affil[1]{\uantof}

\author[2]{J. Zanelli \thanks{E-mail: \href{mailto:z@cecs.cl}{\nolinkurl{z@cecs.cl}}}}
\affil[2]{\cecs}

\maketitle

\begin{abstract}
 We discuss an embedding of $su(n)$ rank-two antisymmetric supercharges in the $su(2,2|d_n)$ superalgebra, where $d_n=n(n-1)/2$. We describe an algorithm to construct the explicit form of the generators of the superalgebra.
\end{abstract}

\tableofcontents

\section{Introduction}\label{intro}  

In $N$-extended conformal supergravity in four dimensions, the supersymmetry algebra is $su(2,2|N)$, which contains the conformal algebra $o(4,2) \sim su(2,2)$, and $u(N)$ as subalgebras \cite{Fradkin:1985am}. Another possible embedding of the conformal algebra is $osp(M|8N)$, where the conformal algebra is embedded as
\begin{equation}
 osp(8N,R) \supset u(2N,2N) \supset su(2,2) \times u(n)\,.
\end{equation}
The possibility of such embeddings comes from the real forms of the corresponding Lie algebras \cite{Satake1960}. These cases appear as extensions of the Haag-Lopuszanski-Sohnius theorem, where the bosonic subalgebra is the direct product of the conformal algebra and an internal symmetry group \cite{Haag:1974qh}. The emergence of $u(n)$ in these two families of superalgebras is generic, meaning that it occurs for the non-centrally extended as well as for the centrally extended algebras \cite{Ferrara:1999si}. In fact the extended algebras could include central charges or charges of rank $p$, on the right hand side of the anticommutators.

The conformal superalgebras are crucial for the construction of extended supergravities \cite{Lauria:2020rhc}, conformal supergravity \cite{Fradkin:1985am,Ferrara:2020zef}, unified theories with matter \cite{Ferrara:1977ij} and superconformal super Yang-Mills theories \cite{Brink:1976bc}. Super Yang-Mills theories have received renewed attention recently due to their rich mathematical structure concerning scattering amplitudes \cite{Arkani-Hamed:2012zlh,Elvang:2013cua}. More recently, these algebras have been also used to construct related theories of matter \cite{Alvarez:2020qmy} and extended MacDowell Mansouri supergravities \cite{Alvarez:2021qbu}.

On an independent line of research we have the grand unified theories with or without supersymmetry that usually take one or more fields transforming in higher rank representations of the gauge group. The archetypal example is the Georgy-Glashow $SU(5)$ model, where many leptons and quarks are included in a $\mathbf{10}$ representation $\chi_{ij}$ \cite{Georgi:1974sy,Langacker:1980js}. An $SU(5)$ model with softly broken supersymmetry was also proposed \cite{Dimopoulos:1981zb}. In order to solve the proton decay and the doublet-triplet splitting problems of the Georgy-Glashow $SU(5)$ model, the flipped $SU(5)$ model was proposed \cite{Barr:1981qv}. Some years later, the supersymmetric flipped $SU(5)$ model that produce hierarchical neutrino masses, was proposed \cite{Antoniadis:1987dx,Ellis:1992zr,Ellis:1993ks,Ellis:1992nq}. Flipped $SU(5)$ can also be embedded in string theory \cite{Antoniadis:1988tt,Campbell:1987gz,Lopez:1992kg}.

Minimal flipped $SU(5)$ GUT was shown to survive experimental electroweak limits \cite{Ellis:2002vk}. Regions of the parameter space that yield proton lifetime estimates which are testable at Hyper-Kamiokande and other next generation experiments have been recently explored \cite{Mehmood:2020irm}. A model that accommodates all constraints, has vanishing cosmological constant and incorporates Starobinsky-like inflation, was proposed in \cite{Ellis:2013nka}. The flipped $SU(5)$ GUT has also been embedded in no-scale supergravity models \cite{Ellis:2017jcp}. Such model can generate no-scale inflation, similar to Starobinsky's inflation, \cite{Ferrara:2013rsa}, and makes predictions for cosmic microwave background observables that can be used to constrain significantly the parameters of the model.

In this paper we describe an embedding of supercharges that transform as the components of a rank-two antisymmetric tensor of $SU(n)$ in the conformal superalgebras $su(2,2|d_n)$, with $d_n=n(n-1)/2$. In section \ref{rank2} we give the basic definitions, explain the embedding of $SU(n)$ generators and the closure of the superalgebra. In section \ref{algo}, we provide an algorithm to obtain the explicit form of generators and structure constants. In section \ref{superalgebra} we provide the explicit form of the superalgebra. In section \ref{final} we summarize our results and provide some comments on special cases.

\section{$su(n)$ rank two supercharges}\label{rank2}   

Let us consider rank two supercharges $\QQ^\alpha_{ij}$. Being antisymmetric, $\QQ^\alpha_{ij}=-\QQ^\alpha_{ji}$, and rank two means that they must commute with $su(n)$ generators, $\TT_I$, by
\begin{equation}\label{Qtrans}
 [\TT_I,\QQ^\alpha_{ij}]=i(t_I)_{ij}{}^{kl}\QQ^\alpha_{kl}\,, \qquad [\TT_I,\QQb_\alpha^{ij}]=-i\QQb_\alpha^{kl} (t_I)_{kl}{}^{ij}\,,
\end{equation}
where $I=1,\cdots,n^2-1$ and $i=1,\cdots,n$. We will consider a complex representation and therefore the generators $\QQb_\alpha^{ij}$ are to be consider as independent from the $\QQ^\alpha_{ij}$ generators and $\alpha=1,\cdots,4$ is a four dimensional spinor index. The coefficients $(t_I)_{ij}{}^{kl}$ carry a rank two antysymmetric hermitian representation of $su(n)$, and are defined by
\begin{equation}\label{rank2gen}
 (t_I)_{ij}{}^{kl}=\Delta^{mn}_{ij}(\lambda_I)_m{}^p \delta^q_n \Delta^{kl}_{pq}\,,
\end{equation}
where the $\lambda_I$ generators belong to the fundamental representation of $su(n)$, see appendix \ref{AppA} for more details and conventions. The commutator of $t_I$ generators is given by
\begin{equation}
 [t_I,t_J]=i f_{IJ}^K \, t_K\,,
\end{equation}
where
\begin{equation}
 [t_I,t_J]_{ij}{}^{kl}\equiv (t_I)_{ij}{}^{pq}(t_J)_{pq}{}^{kl}-(t_J)_{ij}{}^{pq}(t_I)_{pq}{}^{kl}\,.
\end{equation}
See appendix \ref{AppA} for the matrix representation of the $\lambda_I$ generators. In (\ref{rank2gen}), we used the generalized Kronecker delta given by
\begin{equation}
 \Delta^{ij}_{kl}=\frac{1}{2}\left( \delta^i_k\delta^j_l-\delta^i_l\delta^j_k \right)\,,
\end{equation}
where $i=1,\cdots,n$. It has the the property
\begin{equation}
 \Delta^{ij}_{mn}\Delta^{mn}_{kl}=\Delta^{ij}_{kl}\,,
\end{equation}
and
\begin{equation}
 \text{tr}(\Delta)\equiv\Delta^{ij}_{ij}=d_n\equiv\frac{n(n-1)}{2}\,.
\end{equation}
It is worth noting that $t_I$ form an orthogonal set of generators since the trace property is given by
\begin{equation}\label{tracenormalization}
 \text{tr}(t_I t_J)=(t_I)_{ij}{}^{kl} (t_J)_{kl}{}^{ij}=\frac{n-2}{2}\delta_{IJ}\,.
\end{equation}

In a finite matrix representation of the superalgebra we can use
\begin{equation}\label{Qs}
(\QQ^\alpha_{ij})^A_{\ B}=2\Delta^A_{ij} \delta^\alpha_B\,, \qquad
(\QQb_\alpha^{ij})^A_{\ B}=2\delta^A_\alpha \Delta^{ij}_B\,.
\end{equation}
The index $A$ defines a $(4+d_n) \times (4+d_n)$ matrix representation. For this purpose let us choose the unraveling of the index as
\begin{equation}\label{index}
 A=\{ \alpha , [ij] \}\,.
\end{equation}
This unraveling is based on a mapping between the rank two antysymmetric index and an ``unraveled'' index,
\begin{equation}\label{map}
 r:[ij] \xmapsto[]{\qquad } i'\,,
\end{equation}
see first two columns of table \ref{maptable}, see also the appendix \ref{AppB} for more comments on the spacetime range.
\begin{table}[h]
\begin{center}
\begin{tabular}{|c|c||c|}\hline
  $[i,j]$ & $i'$ & $A$ \\
  \hline
   $1,2$ & $1$ & $\alpha+1$ \\
   $1,3$ & $2$ & $\alpha+2$ \\
  $\vdots$ & $\vdots$ & $\vdots$ \\
  $1,n$ & $n-1$  & $\alpha+n-1$ \\
  $2,3$ & $n$ & $\alpha+n$ \\
  $\vdots$ & $\vdots$ & $\vdots$ \\
    $n-1,n$ & $d_n$ & $\alpha+d_n$ \\ \hline
\end{tabular}
  \caption{Mapping between the canonically-ordered pairs of indices $i,j$ to a single index $i'$. The third column show the values of the rows and columns index $A$ of the matrix representation of the superalgebra.}
 \label{maptable}
\end{center}
\end{table}

Certain ambiguity may appear when going from an $i'$-type of component to the two index notation, and for that reason it might be useful to consider the action of $r^{-1}$ as the action of the operator $\Delta^{i'}_{ij}$. For instance
\begin{equation}
 \mathcal{T}_{i'=1} \quad \xmapsto[]{\quad r^{-1}: \ \Delta^{i'}_{ij} \quad} \quad \half \left( \mathcal{T}_{i=1,j=2}-\mathcal{T}_{i=2,j=1}\right)\,.
\end{equation}

Let us adopt the convention of using antihermitian generators in the superalgebra,
\begin{equation}\label{Trank2}
 (\TT_{I})^A_{\ B}=2i(t_I^t)^A{}_B=2i(t_I)_B{}^A\equiv 2i\Delta^A_{kl}(t_I)_{ij}{}^{kl} \Delta^{ij}_B\,,
\end{equation}
with $(\TT_{I})^\dagger=-\TT_{I}$ and
\begin{equation}
 [\TT_I,\TT_J]=f_{IJ}{}^K\TT_K\,,
\end{equation}
where $f_{IJ}{}^K$ are the structure constants of $su(n)$. The use of the transposed generator in the definition (\ref{Trank2}) is so that $(\TT_{I})^A_{\ B}$ produce the $su(n)$ transformations on a field-valued element of the superalgebra $\propto \QQb^{ij} \chi_{ij}+\overline{\chi}^{ij} \QQ_{ij}$, where $\chi_{ij}$ carry a $\mathbf{d}_n$ representation of $su(n)$ (instead of a $\mathbf{d}_n^\ast$), see also \cite{Trigiante:2016mnt}.

The supertrace of a generator $G$ is defined by
\begin{equation}
 \langle G \rangle \equiv \text{tr} (\Gamma G) = (\Gamma) ^A{}_B (G) ^B{}_A\,,
\end{equation}
where
\begin{equation}\label{gradop}
 (\Gamma) ^A{}_B =\delta^A_\alpha \delta^\alpha_B -\half \Delta^A_{ij} \Delta^{ij}_B\,.
\end{equation}
This operator gives the fermionic-bosonic grading
\begin{equation}
 [\Gamma,B]=0\,,\quad \{\Gamma,F \}=0\,,
\end{equation}
where $B$ and $F$ are bosonic and fermionic generators respectively.

A note on the convention of summation in (\ref{gradop}) is in order. When contracted, the summation of the index goes over \emph{all} possible values $i,j$, therefore the coefficient $1/2$ of the second term in (\ref{gradop}) is to prevent double counting. As a matrix $\Gamma$ takes the simple form
\begin{equation}
 \Gamma = \text{diag}(\underbrace{1,\cdots,1}_4,\underbrace{-1,\cdots,-1}_{d_n})\,.
\end{equation}

The generators are supertraceless and their quadratic supertraces are given by
\begin{equation}\label{stTT}
 \langle \TT_I \TT_J \rangle = \frac{n-2}{2}\delta_{IJ}\,.
\end{equation}

The supercharges (\ref{Qs}) are supertraceless and their quadratic supertraces can be computed explicitly as well,
\begin{equation}\label{stQQ}
 \langle \QQ^\alpha_{ij} \QQb^{kl}_\beta\rangle=-\delta^\alpha_\beta \Delta^{kl}_{ij}=-\langle \QQb^{kl}_\beta \QQ^\alpha_{ij}\rangle\,.
\end{equation}

In the next section we will show that extra bosonic generators are required for the closure of the superalgebra.

\subsection{Closure of the superalgebra}

The existence of the mapping $r$ indicates the possibility of embedding the rank two antisymmetric supercharges (\ref{Qs}) in a $sl(4,d_n;\CC)$ superalgebra. In fact their nontrivial anticommutators are
\begin{equation}\label{QQacomm}
 \{ \QQ_{ij}^\alpha,\QQb^{kl}_\beta \}^A{}_B = 4 \delta^\alpha_\beta \Delta^A_{ij} \Delta^{kl}_B+4\Delta^{kl}_{ij}\delta^A_\beta \delta^\alpha_B\,.
\end{equation}
where $\delta^A_\beta \delta^\alpha_B$ spans the $4\times 4$ upper-left block of the matrix representation and $\Delta^A_{ij} \Delta^{kl}_B$ spans the $d_n \times d_n$ lower-right block of the matrix representation:
\begin{equation}
 \left[\begin{array}{c|c}
\delta^A_\beta \delta^\alpha_B &\delta^A_\beta \Delta^{kl}_B\\ [0.5em] \hline
\Delta^A_{ij} \delta^\alpha_B & \Delta^A_{ij} \Delta^{kl}_B\end{array}\right]\,.
\end{equation}
In the upper-right and lower-left blocks we have the supercharges (\ref{Qs}). The right hand side of (\ref{QQacomm}) is supertraceless.

We are interested in two particular real forms of $sl(4,d_n;\CC)$: $su(2,2)$ and $su(m)$ (with $m=d_n$). The generators of $su(2,2)$ can be identified with the generators of the conformal group, that are embedded in the traceless generators of a Clifford algebra. Including the identity matrix, these generators span the second term in the r.h.s. of (\ref{QQacomm}). These generators inherit a reality condition from the $\gamma$-matrices that span the Clifford algebra,
\begin{equation}\label{gamma}
 \gamma_C^\dagger=\varepsilon_C \gamma^0 \gamma_C \gamma^0\,, \qquad \varepsilon_C=\begin{cases}
-1\,, \quad \text{for} \, C=\varnothing\\
+1\,, \quad \text{for} \, C=a\\
+1\,, \quad \text{for} \, C=ab\\
+1\,, \quad \text{for} \, C=\tilde{a}\\
+1\,, \quad \text{for} \, C=5\\
\end{cases}\,.
\end{equation}
where $\gamma_\varnothing=\mathds{1}$, $\gamma_5=i\gamma^0 \gamma^1 \gamma^2 \gamma^3$, $(\gamma_5)^2=\mathds{1}$ and $\gamma_{\tilde{a}}=\frac{i}{3!}\epsilon_{abcd}\gamma^{bcd}=-\gamma_5\gamma_a$.
The $4\times 4$ matrices $\gamma_C$ form a Clifford algebra, $a=1,\cdots,4$ is a spacetime index and $\{\gamma_a,\gamma_b\}=2 \eta_{ab}$, where the metric $\eta$ is given by $\eta=\mathrm{diag}(-,+,+,+)$. The reality condition in the $su(m)$ sector is given by $(\VV_X)^\dagger=-\VV_X$ (anti-hermitian).

The generators $\VV_X$ can be splitted into $\{ \VV_I, \VV_{\tilde{X}} \}$, where $\VV_I$ generate a $SU(n)$ group, however, the set of $\VV_I$ do not generate the right transformations when acting on the rank two supercharges (\ref{Qtrans}). In the next section we will show how the embedding of the $\TT_I$ generators is realized in $su(m)$.

\subsection{$T_I$ embedding}\label{embedding}

It might be illuminating to consider the combinations
\begin{equation}\label{tracelesscomb}
 \delta^A_\beta \delta^\alpha_B-\frac{1}{4}\delta^\alpha_\beta \delta^A_\gamma \delta^\gamma_B \,,
\end{equation}
that are traceless in the $A,B=\alpha,\beta$ block of the matrix representation and have no nontrivial elements in the $A,B=i',j'$ block of the matrix, therefore they are also supertraceless. These combinations span $sl(4)$ that upon imposing the reality condition mentioned above will span the conformal algebra, see appendix \ref{AppB}. By replacing (\ref{tracelesscomb}) in (\ref{QQacomm}) we obtain
\begin{equation}\label{QQacomm2}
 \{ \QQ_{ij}^\alpha,\QQb^{kl}_\beta \}^A{}_B = 4\Delta^{kl}_{ij}(\delta^A_\beta \delta^\alpha_B-\frac{1}{4}\delta^\alpha_\beta \delta^A_\gamma \delta^\gamma_B)+2 \delta^\alpha_\beta (S_{ij}{}^{kl})^A{}_B\,.
\end{equation}
The new generators in the r.h.s. are given by
\begin{equation}
 (S_{ij}{}^{kl})^A{}_B=\frac{1}{4}\delta^A_\alpha \delta^\alpha_B \Delta^{kl}_{ij}+2\Delta^A_{ij}\Delta^{kl}_B\,.
\end{equation}
These generators are supertraceless. The generators of the form $S_{ij}{}^{ij}$ have nontrivial elements in the diagonal only,
\begin{equation}\label{diagonalS}
 S_{ij}{}^{ij}=\text{diag}(\underbrace{1/8,\cdots,1/8}_4,0,\cdots,\underbrace{1/2}_{\text{in place } i'(i,j)},\cdots,0)
\end{equation}
where no summation is implied and $i'(i,j)$ is given by the mapping in table \ref{maptable}. There are $m$ independent generators (\ref{diagonalS}) that will correspond to linear combinations of the $m-1$ diagonal generators in $su(m)$ and one central charge of the superalgebra.

The central charge of the superalgebra will correspond to a linear combinations of (\ref{diagonalS}) that is proportional to the identity in both subspaces $A,B=\alpha,\beta$ or $A,B=[ij],[k,l]$,
\begin{equation}
\sum_{ij} c(i,j) S_{ij}{}^{ij} \propto \text{diag}(\underbrace{1,\cdots,1}_4,\underbrace{1,\cdots,1}_{d_n})
\end{equation}
Clearly there is only one linearly independent combination with such property and can be written as
\begin{equation}\label{Z}
 \ZZ^A_{\ B}=i z \left(\delta^A_\alpha \delta^\alpha_B+\frac{4}{d_n}\Delta^A_{ij}\Delta^{ij}_B\right)\,,
\end{equation}
where $z$ is an arbitrary normalization constant. The generator $\ZZ$ commutes with the bosonic generators, but has nontrivial commutation relations with the supercharges,
\begin{equation}
 [\ZZ,\QQ_{ij}^\alpha]=i z \left(4/d_n-1\right)\QQ_{ij}^\alpha\,, \qquad [\ZZ,\QQb^{ij}_\alpha]=i z \left(4/d_n-1\right)\QQb^{ij}_\alpha\,.
\end{equation}

The generators that are not of the form $S_{ij}{}^{ij}$ (no summation here) span $sl(m)$: they have nontrivial elements in place $A=i'$ and $B=j'$ only, and are traceless in the $sl(m)$ subspace since $j' \ne i'$,
\begin{equation}\label{Soffdiag}
 S_{ij}{}^{kl}=\half \varepsilon_{(i,j)}\varepsilon_{(k,l)}\delta^A_{i'(i,j)}\delta ^{j'(k,l)}_B\,, \quad \text{(no sum in $i,j$ and $p,q$)}.
\end{equation}
where $\varepsilon_{(i,j)}$ is given by
\begin{equation}
 \varepsilon_{(i,j)}=\begin{cases}
+1\,, \quad \text{if $i,j$ is canonically ordered}\\
-1\,, \quad \text{if $i,j$ otherwise,}
                       \end{cases}
\end{equation}
where by canonically ordered we refer to table \ref{maptable}. Expression (\ref{Soffdiag}) is valid when at least one of $i$ or $j$ is not equal to $k$ or $l$.

A key point to note it that after imposing the reality conditions on $S_{ij}{}^{kl}$, we can expand all $su(n)$ generators and the central charge
\begin{align}
 S_{ij}{}^{ij}& \quad \underrightarrow{\quad \text{diagonal generators plus central charge} \quad} \quad \ZZ,\TT_3,\TT_8,\TT_{15},\cdots\\
 S_{ij}{}^{kl}& \quad \underrightarrow{\quad \text{off diagonal generators} \quad} \quad \TT_1,\TT_2,\TT_4,\TT_5,\TT_6,\TT_7,\cdots
\end{align}
Of course the diagonal generators $\TT_3,\TT_8,\TT_{15},\cdots$ correspond to linear combinations of (\ref{diagonalS}) such that the sum has trivial elements in the upper-left block,
\begin{equation}
\TT_{I_\text{diag}} = \sum_{ij} c(i,j) S_{ij}{}^{ij}=\text{diag}(\underbrace{0,\cdots,0}_4,it_{I_\text{diag}}^t)
\end{equation}

The general coefficients of the expansion,
\begin{equation}\label{expan}
 \TT_I=C_I{}^{ij}{}_{kl}S_{ij}{}^{kl}\,,
\end{equation}
can be determined using the property
\begin{equation}
 \langle S_{ij}{}^{kl} S_{pq}{}^{rs} \rangle = -\Delta_{ij}^{rs}\Delta_{pq}^{kl}+\frac{1}{4}\Delta_{ij}^{kl}\Delta_{pq}^{rs}\,,
\end{equation}
therefore
\begin{equation}
-C_I{}^{rs}{}_{pq}+\frac{1}{4}C_I{}^{ij}{}_{ij}\Delta_{pq}^{rs}=\langle  \TT_I S_{pq}{}^{rs} \rangle\,,
\end{equation}
or
\begin{equation}\label{coefs}
 C_I{}^{ij}{}_{kl}=\begin{cases}
-\langle  \TT_I S_{kl}{}^{ij} \rangle\,, \quad \text{for} \quad ij \ne kl\\
-\frac{8}{7}\langle  \TT_I S_{ij}{}^{ij} \rangle\,, \quad \text{for} \quad ij = kl\\
\end{cases}\,.
\end{equation}

By using the coefficients (\ref{coefs}) in (\ref{expan}) we get the explicit form of the embedding $\{\TT_I \} \subset sl(4,m)$ These generators produce the right transformations when acting on the rank two antisymmetric supercharges (\ref{Qtrans}).

In $sl(m,\CC)$ we have $2(m^2-1)$ generators, number that is reduced to $m^2-1$ after imposing the reality condition. In $su(m)$ we have $d_n^2-n^2$ generators that are not in the set of $\{ t_I \}$ generators. Among the missing generators we have $d_n-n$ diagonal generators, and $d_n(d_n-1)-n(n-1)$ off diagonal generators. In the next section we will describe an algorithm to construct the missing generators.

\section{General description of the algorithm}\label{algo} 
\begin{figure}[h]
  \centering
  \includegraphics[trim={0 2 0 6},clip,width=.32\linewidth]{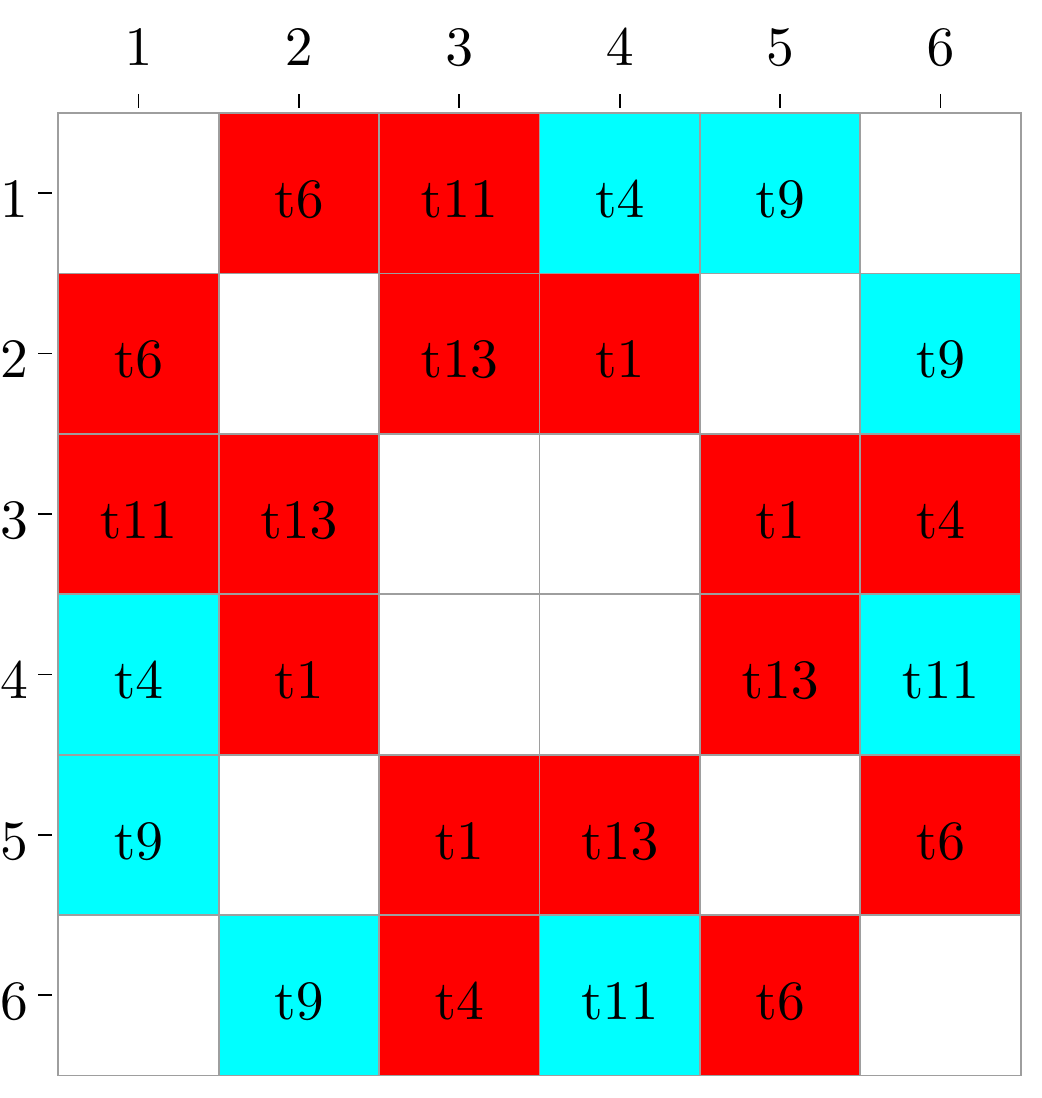}
  \includegraphics[trim={0 2 0 6},clip,width=.32\linewidth]{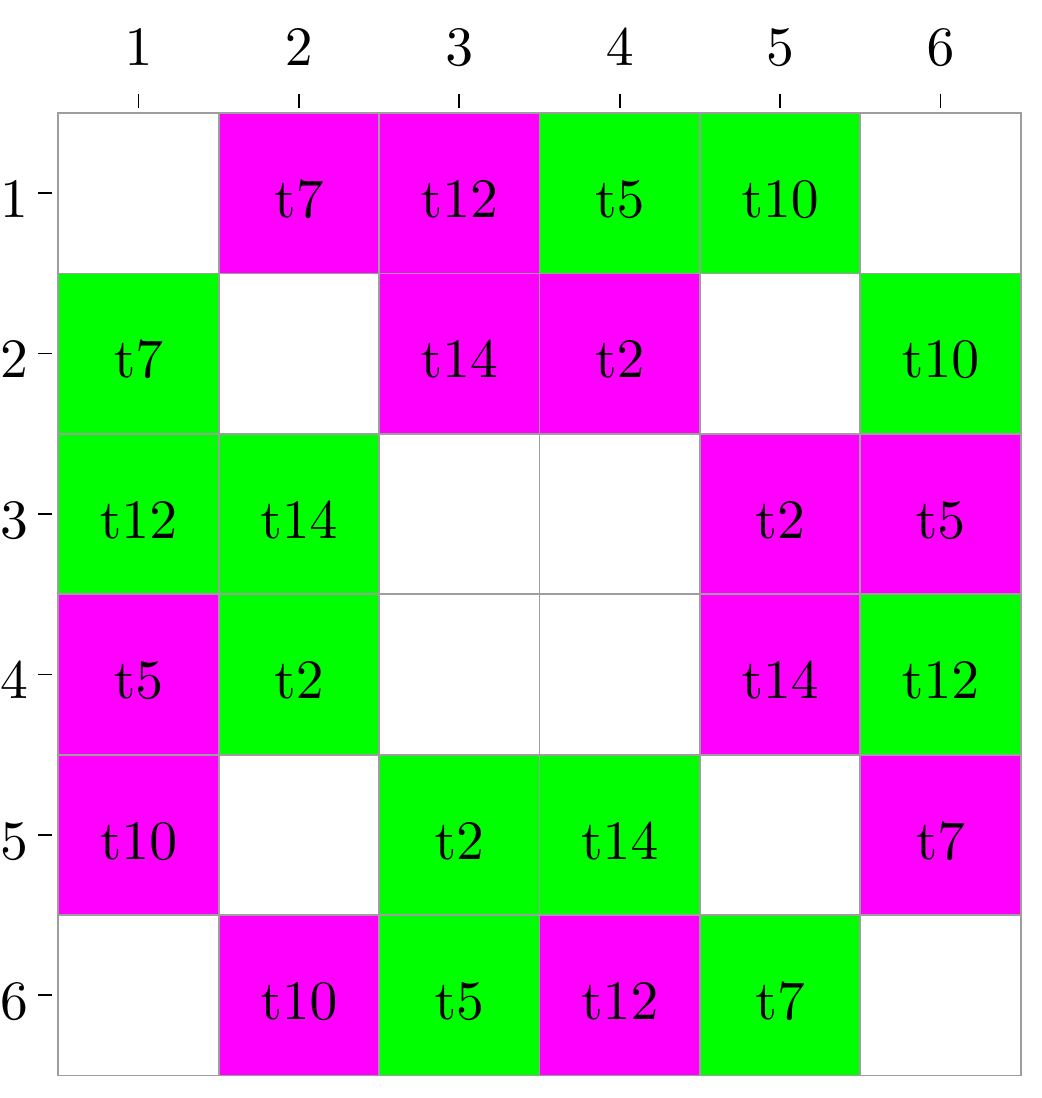}\\
  \includegraphics[trim={0 2 0 6},clip,width=.32\linewidth]{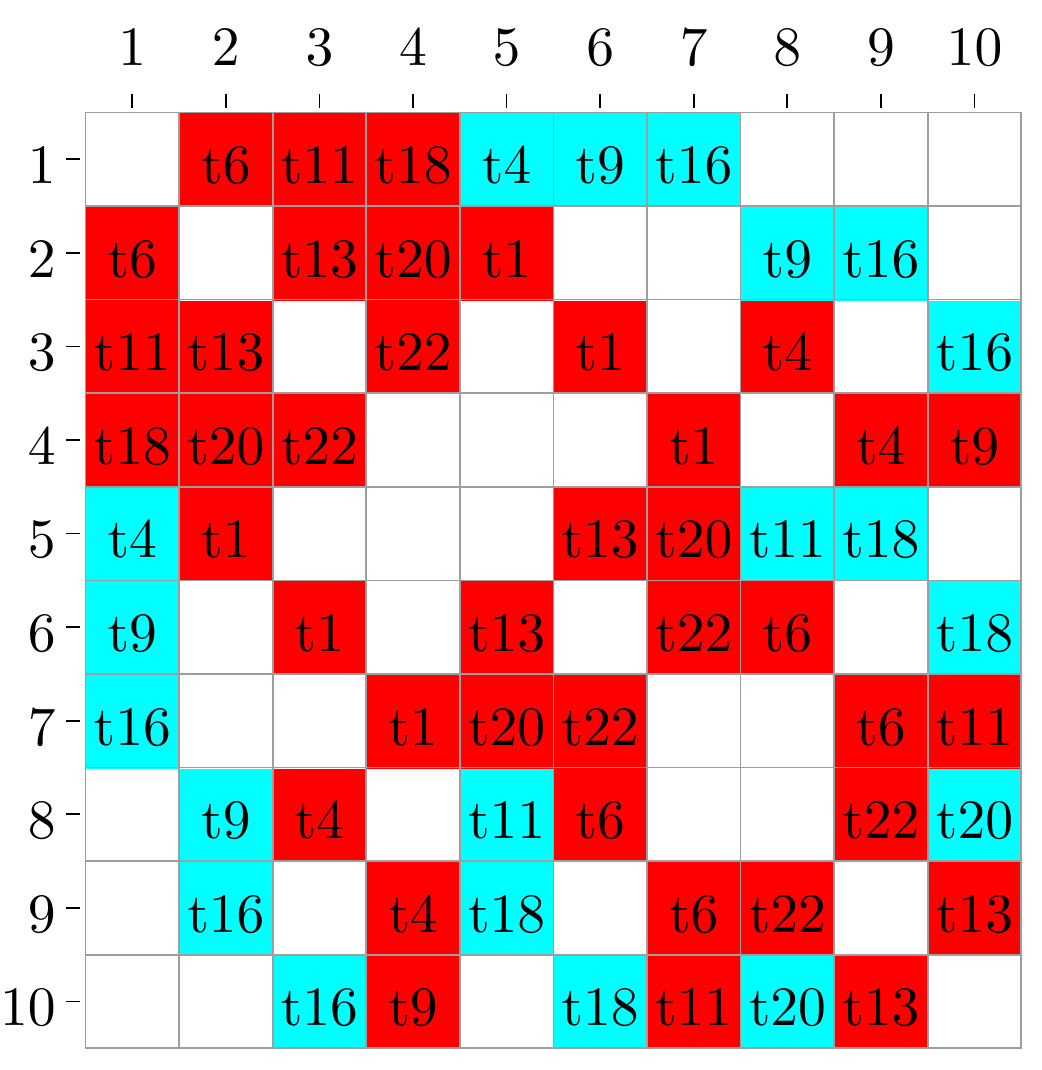}
  \includegraphics[trim={0 2 0 6},clip,width=.32\linewidth]{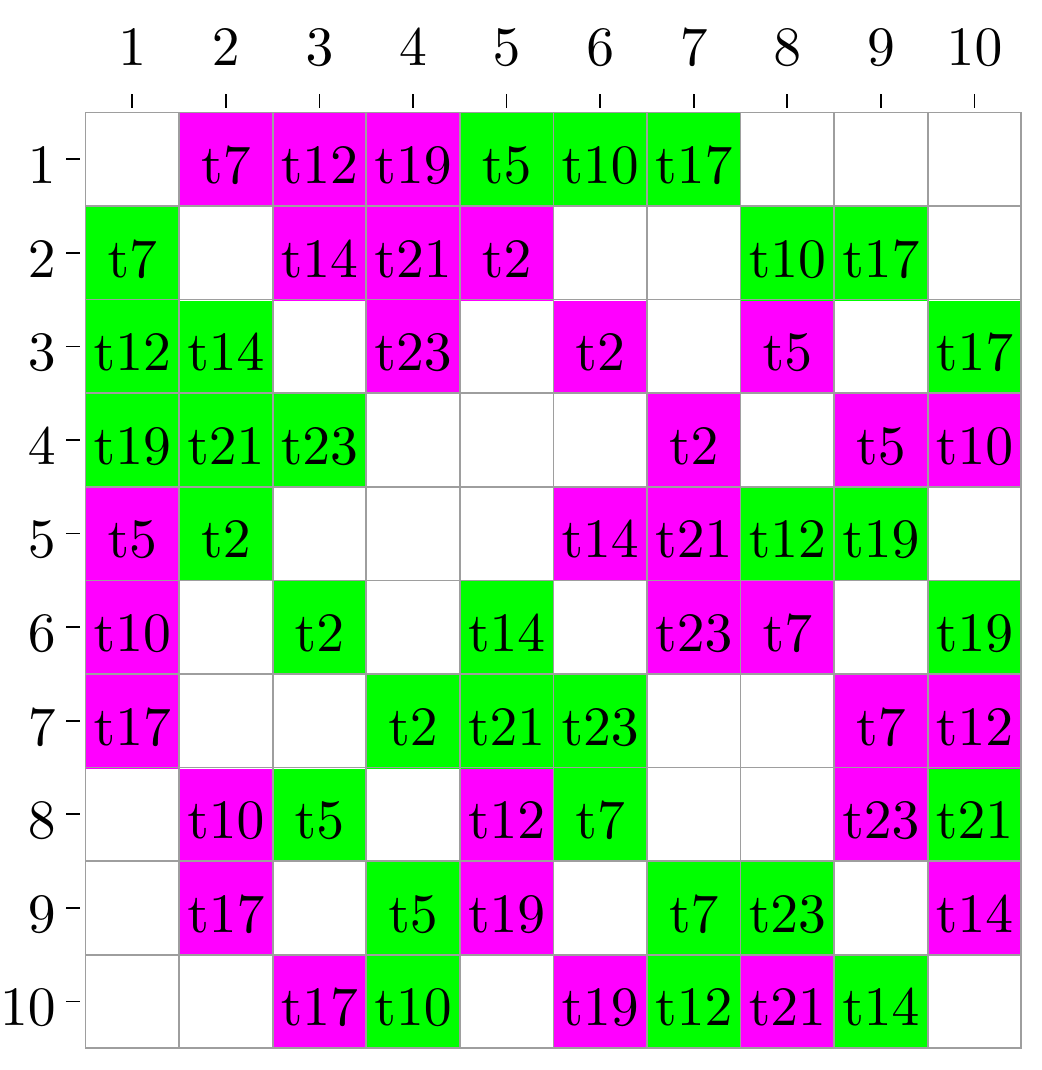}\\
    \includegraphics[width=.5\linewidth]{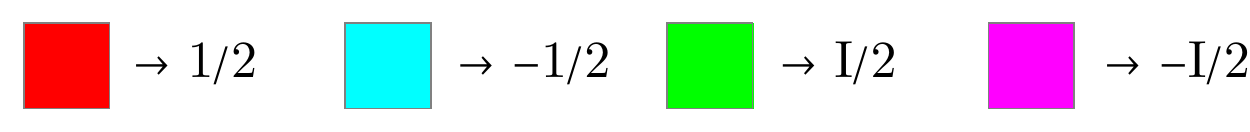}
\caption{Matrix components of the hermitean generators $t_I$ given by (\ref{Trank2}), where the map (\ref{map}) is used to obtain a matrix representation. The off diagonal generators have $2\times (n-2)$ off-diagonal components, as depicted here for $n=4$ and $n=5$. The diagonal generators are given in formulas (\ref{g_3,n=4})-(\ref{g_15,n=4}) for $n=4$, and (\ref{g_3,n=5})-(\ref{g_24,n=5}) for $n=5$.}
\label{tIcompfig}
\end{figure}

Let us start the discussion by looking at the number of nontrivial components of the off-diagonal generators $t_I$. Let us rewrite (\ref{rank2gen}) as follows
\begin{equation}\label{rank2genv2}
 (t_I)_{ij}{}^{kl}=\Delta^{mq}_{ij}(\lambda_I)_m{}^p \Delta^{kl}_{pq}\,.
\end{equation}
From this expression we can see that for each off diagonal generator, $q$ has to take values so that $q\ne m$ and $q\ne p$, and therefore each rank two generator has $(n-2)$ nontrivial components above the diagonal and $(n-2)$ nontrivial components below the diagonal. Figure \ref{tIcompfig} displays the off-diagonal $SU(4)$ and $SU(5)$ generators $t_1, t_2, t_4, t_5, t_6, t_7,..$ by indicating the positions of their non-vanishing entries, while the color indicates the values of those entries. For instance, the upper panels indicates that the $SU(4)$ generator $t_1$ has entries 1/2 at (2,4), (3,5), (4,2) and (5,3); generator $t_9$ has entries -1/2 at (1,5), (2,6), (5,1) and (6,2); etc. We will embed these components in the $d_n \times d_n$ space spanned by the index $i'$ of $su(d_n)$ using the mapping $r$. For the diagonal generators, as an example, let us give explicit expressions for $n=4$,
\begin{align}
    t_3=&\text{diag} (0,\half,\half,\frac{-1}{2},\frac{-1}{2},0)\,,\label{g_3,n=4}\\
    t_8=&\text{diag} (\frac{1}{\sqrt{3}},\frac{-1}{2\sqrt{3}},\frac{1}{2\sqrt{3}},\frac{-1}{2\sqrt{3}},\frac{1}{2\sqrt{3}},\frac{-1}{\sqrt{3}})\,,\\
    t_{15}=&\text{diag} (\frac{1}{\sqrt{6}},\frac{1}{\sqrt{6}},\frac{-1}{\sqrt{6}},\frac{1}{\sqrt{6}},\frac{-1}{\sqrt{6}},\frac{-1}{\sqrt{6}})\,,\label{g_15,n=4}
\end{align}
and $n=5$,
\begin{align}
    t_3=&\text{diag} (0,\half,\half,\half,\frac{-1}{2},\frac{-1}{2},\frac{-1}{2},0,0,0)\,,\label{g_3,n=5}\\
    t_8=&\text{diag} (\frac{1}{\sqrt{3}},\frac{-1}{2\sqrt{3}},\frac{1}{2\sqrt{3}},\frac{1}{2\sqrt{3}},\frac{-1}{2\sqrt{3}},\frac{1}{2\sqrt{3}},\frac{1}{2\sqrt{3}},\frac{-1}{\sqrt{3}},\frac{-1}{\sqrt{3}},0)\,,\\
    t_{15}=&\text{diag} (\frac{1}{\sqrt{6}},\frac{1}{\sqrt{6}},\frac{-1}{\sqrt{6}},\frac{1}{2\sqrt{6}},\frac{1}{\sqrt{6}},\frac{-1}{\sqrt{6}},\frac{1}{2\sqrt{6}},\frac{-1}{\sqrt{6}},\frac{1}{2\sqrt{6}},\frac{-\sqrt{3/2}}{2})\,,\\
    t_{24}=&\text{diag} (\frac{1}{\sqrt{10}},\frac{1}{\sqrt{10}},\frac{1}{\sqrt{10}},\frac{-3}{2\sqrt{10}},\frac{1}{\sqrt{10}},\frac{1}{\sqrt{10}},\frac{-3}{2\sqrt{10}},\frac{1}{\sqrt{10}},\frac{-3}{2\sqrt{10}},\frac{-3}{2\sqrt{10}})\,.\label{g_24,n=5}
\end{align}

Let us denote the generators of $su(d_n)$ in the fundamental representation by $v_X$. We can see that the naive choice $\{t_I,v_{\tilde{X}} \}$, for $\tilde{X}>I$, does not form an orthonormal base of $su(d_n)$ generators, since $\langle t_I,v_{\tilde{X}}\rangle \ne 0$ for some $v_{\tilde{X}}$.

Let us now describe how to construct linearly independent generators that will complete the basis for $su(d_n)$. For the sake of definiteness, let us consider an arbitrary but fixed off diagonal generator $t_I$, see fig. \ref{tIcompfig}. We will use the following list of coefficients,
\begin{align}
 a^{(1)}=&(+1,-1,+1\cdots,+1)\,,\\
 a^{(2)}=&(+1,+1,-1\cdots,+1)\,,\\
 \vdots \nonumber\\
 a^{(n-3)}=&(+1,+1,+1\cdots,-1)\,,
\end{align}
to construct $(n-3)$ new generators that form a linearly independent set with the $t_I$ generator. In order to do so we will multiply each off diagonal $t_I$ generator by a list of coefficients $v_i$ in the following way:
\begin{equation}\label{TLIgen}
 t^{(x)}_I=\left(\begin{array}{c}
      a^{(x)}_1 \times \text{$1^\text{st}$ n.r. of $t_I$ a.d.}\\
      a^{(x)}_2 \times \text{$2^\text{nd}$ n.r. of $t_I$ a.d.}\\
      \vdots\\
      a^{(x)}_{n-2} \times \text{$(n-2)^\text{th}$ n.r. of $t_I$ a.d.}
     \end{array}\right)+
     \left(\begin{array}{c}
      a^{(x)}_1 \times \text{$1^\text{st}$ n.r. of $t_I$ b.d.}\\
      a^{(x)}_2 \times \text{$2^\text{nd}$ n.r. of $t_I$ b.d.}\\
      \vdots\\
      a^{(x)}_{n-2} \times \text{$(n-2)^\text{th}$ n.r. of $t_I$ b.d.}
     \end{array}\right)
\end{equation}
where n.r. stands for ``nontrivial row'', a.d. stands for ``above the diagonal'' and b.d. stands for ``below the diagonal''. We have now $t^{(x)}_I$ $x=1,\cdots,n-3$ generators, that can be added to the original set of off-diagonal generators $t_I$. It is convenient to use the Gram-Schmidt process with the normalization constant given in (\ref{tracenormalization}) to adequately incorporate the full set of diagonal and off-diagonal generators $\{t_I\}$ and the set of $\{t_I^{(x)}\}$ generators. As a result we end up with $(n-1)$ diagonal generators plus $n(n-1)(n-2)$ off diagonal generators that are orthogonal. Similar idea can be used to construct generators $\TT_I^{(x)}$ directly embedded in the superalgebra, where the normalization constant in the Gram-Schmidt process should be given by (\ref{stTT}).

There are also $d_n-n$ diagonal generators and $d_n(d_n-1)-n(n-1)(n-2)$ off-diagonal generators that are in $su(d_n)$ but not yet included so far.

We can complete the off-diagonal generators by adding all the $v_X$ generators that correspond matrix elements that are not fulfilled by the set of $\{ t_I\}$ generators and the previously constructed $\{ t^{(i)}_I\}$ generators, see empty off-diagonal matrix elements in fig. \ref{tIcompfig}. The $v_X$ generators are already orthogonal so there is only a normalization coefficient to adjust in order to comply with the normalization factor in (\ref{tracenormalization}).

Finally, we can add the diagonal generators in $v_{X_\text{diag}}$ with $X_\text{diag} > I_\text{diag}$, by using the Gram-Schmidt process with the normalization constant given in (\ref{tracenormalization}).

After completing the set of $d_n$ generators by the procedure mentioned above, we have a set of generators with matrix elements $(g_{X'})_{i'}{}^{j'}$. We can get the antisymmetric rank two components by using the relation
\begin{equation}\label{rank2g}
 (g_{X'})_{i,j}{}^{k,l}=\half \Delta_{i,j}{}^{p,q}\varepsilon_{(p,q)}\varepsilon_{(r,s)}(g_{X'})_{i'(p,q)}{}^{j'(r,s)}\Delta_{r,s}{}^{k,l}
\end{equation}
where a sum $\sum_{p,q}\sum_{r,s}$ is implied. Let us note that functional form of (\ref{rank2g}) should not be confused with the functional form of (\ref{rank2genv2}), even though in terms of components is valid that
\begin{equation}
 (g_I)_{ij}{}^{kl}=(t_I)_{ij}{}^{kl}\,.
\end{equation}
Also let us make clear that when the double sum in $p,q$ and $r,s$ is used to construct the trace-like operation there is an extra factor because of the double counting,
\begin{equation}
 \sum_{p,q}\sum_{r,s}(g_{X'})_{i'(p,q)}{}^{j'(r,s)}(g_{Y'})_{j'(p,q)}{}^{i'(r,s)}=4 \times \frac{n-2}{2}\delta_{X'Y'}\,.
\end{equation}

As a final result we have a set of generators $g_{X'}$ such that
\begin{equation}\label{tracenormalizationg}
 \text{tr}(g_{X'} g_{Y'})=\frac{n-2}{2}\delta_{X'Y'}\,.
\end{equation}
The generators $\{g_{X'}\}$ split naturally as $\{g_{X'}\}=\{g_I,g_{\tilde{X}}\}$ where $\tilde{X}>n^2-1$ and $g_I=t_I$.

The generators embedded in the superalgebra are given by the generalization of (\ref{Trank2})
\begin{equation}\label{GAB}
 (\GG_{X'})^A_{\ B}=2i(g_{X'}^t)^A{}_B=2i(g_{X'})_B{}^A\equiv 2i\Delta^A_{kl}(g_{X'})_{ij}{}^{kl} \Delta^{ij}_B\,.
\end{equation}

\subsection{Step by step algorithm} 

Let us summarize the algorithm described in this section by the following list:
\begin{enumerate}
 \item Take the fundamental representation of $su(n)$.
 \item Construct rank two $su(n)$ generators using (\ref{Trank2}), (\ref{rank2gen}).
 \item Construct rank two supercharges using (\ref{Qs}).
 \item Map the generators of the two previous steps to a $(4+d_n)\times (4+d_n)$matrix representation using (\ref{index}) and (\ref{map}).
 \item Construct the central charge by using (\ref{Z}).
 \item Construct L.I. generators $\TT_I^{(x)}$, $x=1,\cdots,n-3$ associated to every off-diagonal $su(n)$ generator, use (\ref{TLIgen}) to this purpose. In this step and the following two steps the normalization constant in the Gram-Schmidt process should be given by (\ref{stTT}).
 \item Add off-diagonal $su(d_n)$ generators embedded in the superalgebra $\VV_X=i v_X$ (that correspond to empty sites in figure \label{tIcomp}) and to the set of generators using the Gram-Schmidt process.
 \item Include the diagonal $su(d_n)$ generators embedded in the superalgebra $\VV_{X_\text{diag}}$ with $X_\text{diag} > I_\text{diag}$, by using the Gram-Schmidt process.
\end{enumerate}

\section{Superalgebra}\label{superalgebra} 

In this section we will give a summary with the explicit form of the algebra.

In order to write the anticommutators of the superalgebra in terms of $\TT_I$ we have two options: (i) invert the relation (\ref{expan}) and replace $S_{ij}{}^{kl}$ in (\ref{QQacomm2}); or (ii) use the standard form of the $su(2,2,d_n)$ superalgebra, replace the expression (\ref{inverserelation}) where it corresponds and use the inverse map $r^{-1}: \ i' \xmapsto{ } i,j$. Either way the final superalgebra takes the form
\begin{eqnarray}
&[\GG_{X'},\QQ^\alpha_{ij}]=i(g_{X'})_{ij}{}^{kl}\QQ^\alpha_{kl}\,, \quad [\GG_{X'},\QQb_\alpha^{ij}]=-i\QQb_\alpha^{kl} (g_{X'})_{kl}{}^{ij}\,,&\\
&[\ZZ,\QQ^\alpha_{ij}]=iz\left(4/d-1\right)\QQ^\alpha_{ij}\,, \quad [\ZZ,\QQb_\alpha^{ij}]=-iz\left(4/d-1\right)\QQb_\alpha^{ij}\,,&\label{Zcomm}\\
&\{\QQ^\alpha_{ij},\QQb_\beta^{kl}\}=(\gamma^C)^\alpha{}_\beta \JJ_C-\left(\frac{4i}{n-2}(g_{X'})_{ij}{}^{kl}\GG_{X'}+\frac{i}{2z}\Delta_{ij}{}^{kl} \ZZ\right)\delta^{\alpha}_{\beta}\,.&
\end{eqnarray}
where
\begin{equation}
(\gamma^C)^\alpha{}_\beta \JJ_C=4\Delta^{kl}_{ij}  (\delta^A_\beta \delta^\alpha_B-\frac{1}{4}\delta^\alpha_\beta \delta^A_\gamma \delta^\gamma_B )\,,
\end{equation}
The generators $\JJ_C$ span the conformal group, see appendix \ref{AppB}.

The commutators between $\GG_{X'}$ generators is given by
\begin{equation}
 [\GG_{X'},\GG_{Y'}]= f_{X'Y'}{}^{Z'} \GG_{Z'}\,,
\end{equation}
where the structure constants are given below.

The quadratic supertraces are given by (\ref{stQQ}) and
\begin{equation}
 \langle \GG_{X'} \GG_{Y'} \rangle = \frac{n-2}{2}\delta_{X'Y'}\,, \qquad \langle \ZZ^2 \rangle = 4z^2 (4/d_n-1)\,.
\end{equation}

\subsection{$su(m)$ structure constants in the $g_{X'}$ basis}

The generators $g_{X'}$ are related to $v_X$ by means of the change of basis matrix defined by
\begin{equation}
 g_{X'} = C_{X'}{}^X v_X\,.
\end{equation}
We will define the inverse matrix by
\begin{equation}\label{inverserelation}
 v_X = C_{X}{}^{X'} g_{X'}\,.
\end{equation}
It satisfies
\begin{equation}
 C_{X}{}^{X'}C_{X}{}^{Y'}=\frac{1}{n-2}\delta^{X'Y'}\,.
\end{equation}

Let us define the structure constants of $su(m)$ in the fundamental representation by
\begin{equation}
 [v_X,v_Y]=i F_{XY}{}^Z v_Z\,.
\end{equation}
Using the matrix $C_{X}{}^{X'}$ and its inverse we can determine the structure constants $f_{X'Y'}{}^{Z'}$, that are defined in the $g_{X'}$ basis
\begin{equation}
 [g_{X'},g_{Y'}]=i f_{X'Y'}{}^{Z'} g_{Z'}\,,
\end{equation}
where
\begin{equation}
f_{X'Y'}{}^{Z'} =C_{X'}{}^X C_{Y'}{}^Y F_{XY}{}^Z C_{Z}{}^{Z'}\,.
\end{equation}
The following relations are satisfied for any $n$
\begin{equation}
  f_{X'Y'}{}^{Z'}=\begin{cases}
          f_{IJ}{}^{K}=F_{IJ}{}^{K}\,, \quad \text{if } X'=I,\ Y'=J,\ Z'=K  \\
          f_{IJ}{}^{\widetilde{X}}=0\,, \quad \text{if }  X'=I,\ Y'=J,\ Z'=\widetilde{X}\,.
        \end{cases}
\end{equation}
Of course $f_{\widetilde{X}\widetilde{Y}}{}^{I}\ne 0$ because $su(m)$ is simple. Generically $f_{\widetilde{X}\widetilde{Y}}{}^{\widetilde{Z}}\ne0$ except for $n=4$ where $f_{\widetilde{X}\widetilde{Y}}{}^{\widetilde{Z}}=0$.

\section{Conclusions}\label{final} 

The superalgebra obtained in this paper is isomorphic to $su(2,2|d_n)$, but it is written in a base such that the supercharges transform as rank two tensors under the $su(n)$ generators $\TT_I=\GG_I \in su(2,2|d_n)$.

As special cases appear $n=3$, where the rank two algebra is isomorphic to $su(n)$. Less trivial is the case $n=4$, where the structure constants $f_{\widetilde{X}\widetilde{Y}}{}^{\widetilde{Z}}=0$, where $\widetilde{X},\widetilde{Y},\widetilde{Z} > I$. It is interesting to explore if this possibility has physical realizations and its consequences for spontaneous symmetry breaking. Note that the special case $n=4$ does not correspond to the abelian central charge, (\ref{Zcomm}). In fact it is interesting to study if there exist physical realizations of the algebra that can circumvent technical difficulties encountered in gauged conformal supergravity models such as \cite{Ferrara:1977ij}.

We adjunct files with the tensors and structure constants for $n=4$ and $n=5$ in the ArXiV repository as sparse arrays. The numerical values of the array elements are listed in the form
\begin{equation}
    \{ i_1, i_2,\cdots , i_l \} \longrightarrow \text{``value''}\,,
\end{equation}
that corresponds to the assignment $T_{i_1, i_2,\cdots , i_l} = \text{``value''}$, see table \ref{filenames}

\begin{table}[h]
\begin{center}
\begin{tabular}{c|c}
  object & file name descriptor \\
  \hline
   $C_{X'}{}^X$ & ``coefdXpuX''\\
   $C_{X}{}^{X'}$ & ``coefdXuXp''\\
   $F_{XY}{}^Z$ & ``FdXdXuX''\\
   $f_{X'Y'}{}^{Z'}$ & ``fdXpdXpuXp''\\
   $(g_{X'})_{ij}{}^{kl}$ & ``gdXpdidiuiui''\\
   $(g_{X'})_{i'}{}^{j'}$ & ``gdXpdipuip''\\
   $(G_{X'})_{i'}{}^{j'}$ & ``GdXpdipuip''\\
   $(G_{X'})^A{}_B$ & ``GdXpuAdA''\\
   map $r$ & ``rArrules''\\
   $(S_{ij}{}^{kl})^A{}_B$ & ``ShatdidiuiuiuAdA''\\
   $(T_I)^A{}_B$ & ``Trank2dIuAdA''\\
   $(v_X)_{ij}{}^{kl}$ & ``vdXdidiuiui'', \\
   $(v_X)_{i'}{}^{j'}$ & ``vdXdipuip''\\
   $(V_X)_{i'}{}^{j'}$ & ``VdXdipuip''\\
   $(V_X)^A{}_B$ & ``dXuAdA''
\end{tabular}
  \caption{Supplemental material: plain text files available in ArXiV.}
 \label{filenames}
\end{center}
\end{table}

\section*{Acknowlegements}

P. A. acknowledges MINEDUC-UA project ANT 1755 and Semillero de Investigación project SEM18-02 from Universidad de Antofagasta, Chile.

\begin{appendices}

\section{Fundamental representation of $su(n)$}\label{AppA} 
In this section we will include the basic formulas and notations related to the fundamental representation of $su(n)$, for more details see e.g., ref. \cite{Pfeifer2003TheLA}. Let us define
\begin{align}
 & L_{i,j}=\half (\delta_{i,j}+\delta_{j,i})\,,\\
 & M_{i,j}=\frac{i}{2} (\delta_{i,j}-\delta_{j,i})\,,\label{num}\\
 & D_k=\frac{1}{2k(k+1)}\left(\sum_{l=1,\cdots,k}\delta_{l,l}-k\delta_{k+1,k+1}\right)\,.
\end{align}
where $i=\sqrt{-1}$ in (\ref{num}) and $\delta_{i,j}$ is the Kronecker delta.

We will contruct the generators of the fundamental representation of $su(n)$ as depicted in the following diagram
\begin{equation}
 \begin{array}{|c|c|c|c|c}\hline
     & j=2 & j=3 & j=4 & \cdots \\ \hline
i=1 & \makecell{\lambda_1=2L_{1,2} \\ \lambda_2=2M_{1,2}} & \makecell{\lambda_4=2L_{1,3}\\ \lambda_5=2M_{1,3}} & \makecell{\lambda_9=2L_{1,4} \\\lambda_{10}=2M_{1,4}} & \\ \hline
i=2 & \lambda_3=2D_1 & \makecell{\lambda_6=2L_{2,3} \\ \lambda_7=M_{2,3}} & \makecell{\lambda_{11}=L_{2,4} \\ \lambda_{12}=2M_{2,4}}&\\ \hline
i=3 &  & \lambda_8=2D_2 & \makecell{\lambda_{13}=L_{3,4} \\ \lambda_{14}=2M_{3,4}}&\\ \hline
i=4 &  &  & \lambda_{15}=2D_3 &\\ \hline
\vdots&&&&\ddots
 \end{array}
\end{equation}

These generators have the following commutation relations
\begin{equation}
 [\lambda_I,\lambda_J]=2i f^{IJK} \lambda_K\,.
\end{equation}
where the structure constants are defined by
\begin{equation}
 f_{IJK}=-\frac{i}{4} \text{tr} (\lambda_I [\lambda_J , \lambda_K])\,.
\end{equation}
The trace property is given by
\begin{equation}
 \text{tr}(\lambda_I \lambda_J)\equiv(\lambda_I)_i{}^j (\lambda_J)_j{}^i=2\delta_{IJ}\,.
\end{equation}
Therefore the $\lambda_I$ generators are a base for $n \times n$ hermitian traceless matrices. The $\lambda$-matrices are also endomorphisms when they act on spinors carrying a $\mathbf{n}$ representation of $su(n)$ as
\begin{equation}
 \psi^\alpha_i \xmapsto{\quad \lambda_I \quad} (\lambda_I)_i^{\ j}\psi^\alpha_j \,.
\end{equation}

\section{Conformal algebra}\label{AppB} 

The set of conformal generators is given by $\{ \JJ_C \} = \{ \JJ_a,\JJ_{ab},\KK_a,\DD \}$ where the embedding in the superalgebra is the following
\begin{align}
&\JJ_a =\left[\begin{array}{c|c}
\frac{s}{2}\gamma_a &  0_{4\times d_n}\\[0.5em] \hline
0_{d_n\times4} & 0_{d_n\times d_n} \label{Ja}\\
\end{array}\right]\,, \quad \text{or} \quad (\JJ_a)^A_{\ B}=\frac{s}{2}(\gamma_a)^\alpha_{\ \beta}\delta^A_{\ \alpha} \delta^\beta_{\ B}=\frac{s}{2}(\gamma_a)^A_{\ B}\,,&\\
&\JJ_{ab} =\left[\begin{array}{c|c}
\frac{1}{4}[\gamma_a,\gamma_b] &  0_{4\times d_n}\\[0.5em] \hline
0_{d_n\times4} & 0_{d_n\times d_n} \\
\end{array}\right]\,, \quad \text{or} \quad (\JJ_{ab})^A_{\ B}=\frac{1}{4}[\gamma_a,\gamma_b]^A_{\ B}=(\Sigma_{ab})^A_{\ B}\,,&\\
&\KK_a =\left[\begin{array}{c|c}
\frac{1}{2}\tilde{\gamma}_a &  0_{4\times d_n}\\[0.5em] \hline
0_{d_n\times4} & 0_{d_n\times d_n} \\
\end{array}\right]\,, \quad \text{or} \quad (\KK_a)^A_{\ B}=\frac{1}{2}(\tilde{\gamma}_a)^A_{\ B}\,,&\\
&\DD =\left[\begin{array}{c|c}
\frac{1}{2}\gamma_5 &  0_{4\times d_n}\\[0.5em] \hline
0_{d_n\times4} & 0_{d_n\times d_n} \\
\end{array}\right]\,, \quad \text{or} \quad (\DD)^A_{\ B}=\frac{1}{2}(\gamma_5)^A_{\ B}\,,&\label{D}
\end{align}
where $\Sigma_{ab}=(1/4)[\gamma_a,\gamma_b]$. Matrices $\gamma_a$, $\gamma_5$ and $\tilde{\gamma_a}$ are defined below (\ref{gamma}). In the second expressions in (\ref{Ja})-(\ref{D}) we have used the convention given in (\ref{index}). This means that all the possible products that mix spaces like $p^{ij}_{\ A}  q^A_{\ \alpha}$, where $p \in su(m)$ and $q \in \text{Clifford}$,  are trivial.
Thus, the following relations are understood
\begin{eqnarray}
&(p)^A_{\ B}=\delta^A_\alpha (p)^\alpha_{\ \beta} \delta^\beta_B\,,&\\
&(q)^A_{\ B}=\Delta^{A}_{kl}(q)^{kl}{}_{ij} \Delta^{ij}_{B}\,.&
\end{eqnarray}

The generators $\JJ_a$ and $\JJ_{ab}$ close in an adS$_4$ algebra,
\begin{equation}
[\JJ_a,\JJ_b]=s^2J_{ab}\,,
\end{equation}
\begin{equation}
[\JJ_a,\JJ_{bc}]=\eta_{ab}J_c-\eta_{ac}J_b\,,
\end{equation}
\begin{equation}
[\JJ_{ab},\JJ_{cd}]=-(\eta_{ac}\JJ_{bd}-\eta_{ad}\JJ_{bc}-\eta_{bc}\JJ_{ad}+\eta_{bd}\JJ_{ac})\,.
\end{equation}
The parameter $s^2$ can take values $s^2=+1,-1$ for Anti de Sitter or de Sitter algebras respectively. Including $\KK_a$ and $\DD$ they close in the conformal algebra,
\begin{equation}
[\KK_a,\KK_b]=-J_{ab}\,.
\end{equation}
\begin{equation}
[\JJ_a,\KK_{b}]=s\eta_{ab}D\,.
\end{equation}
\begin{equation}
[\KK_a,\JJ_{bc}]=\eta_{ab}\KK_c-\eta_{ac}\KK_b\,.
\end{equation}
\begin{equation}
[\DD,\JJ_{a}]=-s\KK_a\,.
\end{equation}
\begin{equation}
[\DD,\KK_{a}]=-s^{-1}\JJ_a\,.
\end{equation}

The supercharges carry a spin $1/2$ representation of the conformal generators,
\begin{eqnarray}
&[\JJ_a,\QQ^\alpha_{ij}]=-\frac{s}{2}(\gamma_a)^\alpha_{\ \beta}\QQ^\beta_{ij}\,, \quad [\JJ_a,\QQb_\alpha^{ij}]=\frac{s}{2}\QQb_\beta^{ij}(\gamma_a)^\beta_{\ \alpha}\,,&\\
&[\JJ_{ab},\QQ^\alpha_{ij}]=-(\Sigma_{ab})^\alpha_{\ \beta}\QQ^\beta_{ij}\,, \quad [\JJ_{ab},\QQb_\alpha^{ij}]=\QQb_\beta^{ij}(\Sigma_{ab})^\beta_{\ \alpha}\,,&\\
&[\KK_a,\QQ^\alpha_{ij}]=-\frac{1}{2}(\tilde{\gamma}_a)^\alpha_{\ \beta}\QQ^\beta_{ij}\,, \quad [\KK_a,\QQb_\alpha^{ij}]=\frac{1}{2}\QQb_\beta^{ij}(\tilde{\gamma}_a)^\beta_{\ \alpha}\,,&\\
&[\DD,\QQ^\alpha_{ij}]=-\frac{1}{2}(\gamma_5)^\alpha_{\ \beta}\QQ^\beta_{ij}\,, \quad [\DD,\QQb_\alpha^{ij}]=\frac{1}{2}\QQb_\beta^{ij}(\gamma_5)^\beta_{\ \alpha}\,.&
\end{eqnarray}

\end{appendices}

\bibliographystyle{ieeetr}
\bibliography{draft_v1.bib}

\end{document}